\documentclass[preprint,showpacs,preprintnumbers,amsmath,amssymb,aps]{revtex4}

\usepackage{graphicx}
\usepackage{dcolumn}
\usepackage{bm}
\usepackage{natbib}
\usepackage{braket}
\usepackage{units}
\usepackage{amsmath}
\usepackage{nicefrac}
\usepackage[nooneline]{subfigure}

\renewcommand{\d}{ \text{d}}
\newcommand{\e}{ \text{e}}

\begin{document}

\preprint{DO-TH-12/41}

\title{Neutrino-induced pion production at low energies and in the small $Q^2$ region}

\author{E. A. Paschos}
 \email{paschos@physik.uni-dortmund.de}
\author{Dario Schalla}
 \email{dario.schalla@tu-dortmund.de}
\affiliation{Department of Physics, TU Dortmund, D-44221 Dortmund, Germany}

\date{February 25, 2013}

\begin{abstract}
We analyse neutrino-induced reactions in the small $Q^2$ region and for energies covering the production and decay of the delta resonance. One of our results is the agreement with the MiniBooNE data for $1\pi^+$ and $1\pi^0$ final states. In addition we present differential cross sections for charged and neutral currents and for proton and neutron targets. Finally, we present cross sections induced by muon and electron type neutrinos, where effects of the lepton masses are visible.
\end{abstract}


\maketitle

\section{Introduction}
\label{sec:Introduction}
Neutrino interactions in the low and middle energy region are attracting considerable attention because they investigate new properties~\cite{An:2012eh,Ahn:2012nd} and may lead to new discoveries. For this program precise estimates in the low and intermediate energy regions are necessary. 
Over the past few years many articles have been published, sometimes with conflicting results. The differences among the theoretical conclusions and some of the resolutions have been summarized in a recent review article~\cite{Morfin:2012kn}. We mention three issues relevant to our work.
\begin{enumerate}
 \item There were two values for the dominant axial matrix element for the nucleon-$\Delta$ transition. The PCAC prediction~\cite{Lalakulich:2005cs,Paschos:2003qr} with the value $C_5^A(0)=1.2$ and a smaller value of $0.87$ from a fit of the ANL data~\cite{Hernandez:2007qq} with a large nonresonant background. After several studies which modified the background, the results of simultaneous fits for the ANL and BNL data give a larger value $C_5^A(0)=1.10 \pm 0.08$, which is now accepted and prefered~\cite{Graczyk:2009qm,Hernandez:2010bx,Paschos:2011ye}.
 \item A sizable background under the delta resonance was introduced based on diagrams from chilal symmetry~\cite{Hernandez:2007qq,Lalakulich:2010ss}. Electroproduction data~\cite{Galster:1972rh} gives a small background. Furthermore, application of PCAC relates matrix elements of the axial current to pion-proton and pion-neutron data where a nonresonant background is very small.
 \item For experiments on nuclear targets a generator for the transport model must be introduced. The GiBUU code has difficulties in reproducing the MiniBooNE data~\cite{Lalakulich:2011eh,Lalakulich:2011ne,Lalakulich:2012cj,Mosel:2012kt}. Their predictions lie far below the experimental data.
\end{enumerate}

What we need now are reliable estimates and calculations. For this reason we went back to our earlier calculations~\cite{Paschos:2011ye} and selected the low $Q^2$ region as a reliable benchmark for estimating cross sections. In this region the axial contribution is related by PCAC to cross sections and the vector contribution by CVC to data again. The introduction of experimental data includes nonresonant contributions.

As another example of a discrepancy and its resolution, we mention the peak  and the turn over in the differential cross section $\frac{\d \sigma}{\d Q^2}$ as $Q^2 \rightarrow 0$. We found that in addition to the vanishing of form factors at the boundary of the kinematic region, the axial contribution has a peak at $Q^2 \approx \unit[0.03]{GeV^2}$, reflecting the resonance peak in the integration over $W$. An earlier result for the differential cross section was presented and is slightly improved in our figure~\ref{fig:1}, where the cross section is shown as the sum of various contributions. In  figure~\ref{fig:1} the prominence of the peak for the axial contribution is evident. A second property in this figure is the fact, that the vector-squared and the interference contributions are close to each other and add up for $\nu_\mu p  \rightarrow \mu^- \Delta^{++}$. For the antineutrino reaction $\overline{\nu}_\mu n \rightarrow \mu^+ \Delta^-$ they cancel each other leaving the axial current contribution as the dominant one, with the smaller terms also included in our calculations. This introduces a new test for PCAC and, if confirmed, will give a reliable cross section for antineutrinos. The general trend is qualitatively confirmed by experimental results where the antineutrino reactions are always smaller~\cite{Bolognese:1979gf,Barish:1979ny,Grabosch:1988gw}.

In this article we will extend the method to other reactions and will compare them with experimental data when available. We present the differential cross sections $\frac{\d \sigma}{\d Q^2}$ and we shall compare them with MiniBooNE results~\cite{AguilarArevalo:2010bm,AguilarArevalo:2010xt} including nuclear corrections which are explained below. Since the method, including nuclear corrections, is transparent we can identify each contribution explicitly. This gives, in most cases, consistent results and whenever we encounter differences we discuss the modifications that may be necessary. The method of concentrating in the low $Q^2$ region has been adopted by another group~\cite{Kamano:2012id} which proposes to extend it to higher resonances by including a model with coupled channels for the production.

\noindent Two features are special in neutrino reactions:
\begin{enumerate}
\item the possibility to use various types of neutrino beams $\nu_e, \, \nu_\mu, \, \nu_\tau$, and
\item the necessity to have various nuclear targets.
\end{enumerate}
Both bring peculiarities which we will address.
The various neutrino beams will be present in long-baseline experiments because starting with a $\nu_\mu$ beam a $\nu_e$ component will be generated through oscillations. In the kinematic region of this article the variable $Q^2$ is of the same magnitude as $m_\mu^2$ and $m_\pi^2$. We use formulas with exact masses and include all helicities for the lepton vertex. A difference is expected between $\nu_\mu$ and $\nu_e$ reactions which we calculate and present in section~\ref{sec:DescriptionOfTheMethod}.

For nuclear corrections we use a model inspired by Fermi~\cite{Fermi:unpublished,Adler:1974qu}, which is based on an analytic solution of the transport equation with the results summarized in a simple way, so that the various steps are evident. This allows one to identify the origin of the effects. The direct inputs for the initial cross sections combined with the rescattering model brings a rather good agreement with the results of the MiniBooNE experiment~\cite{AguilarArevalo:2010bm,AguilarArevalo:2010xt}. 

An earlier comparison~\cite{Paschos:2011ye} with the Argonne and Brookhaven experiments, in this limited kinematic region, was consistent. The early experiments have limited statistics for the $\nu_\mu n \rightarrow \mu^- \Delta^+$ channels. For this reason the determination of the nonresonant amplitude with $I=\nicefrac{1}{2}$ is at best very approximate. Electroproduction experiments have a small $A^{\nicefrac{1}{2}}$ amplitude~\cite{Galster:1972rh}. Furthermore the reactions $\pi^\pm p \rightarrow \pi^\pm p$ and the charge exchange $\pi^- p \rightarrow \pi^0 n$ are dominated, in this energy range, by the delta resonance, allowing only a small background~\cite{Beringer:1900zz,Sadler:2004yq}. 

All results support the view that our approach is a good first order approximation. The fact that the nuclear corrections bring the theoretical curves closer to the data is well understood and gratifying. 

In section~\ref{sec:DescriptionOfTheMethod} of this article we describe the method and present results for charged and neutral current reactions. In section~\ref{sec:ComparisonWithRecentExperiments} we apply the method to the MiniBooNE results and find good agreement. In the last section we present our conclusions.

\section{Description of the method}
\label{sec:DescriptionOfTheMethod}
We consider the reactions listed in table~\ref{tab:ISOreactions}. The reaction in the first row for $\Delta^{++}$ production gives the larger cross section and has been studied most extensively. The Clebsch-Gordan coefficients (CGC) in the fourth column refer to the $A^{\nicefrac{3}{2}}$ amplitude which is dominant and should give the prominent features of the reactions. In this article, however, we calculate each amplitude separately by using the following method. For the axial current alone we use the cross section
\begin{align}
\frac{\d \sigma^{(A)}}{\d Q^2 \d\nu} = \frac{G_F^2|V_{ud}|^2}{8\pi^2} \frac{\nu f_\pi^2}{E_\nu^2 Q^2} \left[ \tilde{L}_{00} + 2 \tilde{L}_{l0} \frac{m_\pi^2}{Q^2+ m_\pi^2} + \tilde{L}_{ll} \left( \frac{m_\pi^2}{Q^2+ m_\pi^2} \right)^2 \right] \sigma ( \pi^+ p \rightarrow \pi^+ p)
.\end{align}
The matrix elements $\tilde{L}_{00}, \, \tilde{L}_{l0}$ and $\tilde{L}_{ll}$ were introduced in references~\cite{Paschos:2005km,Paschos:2009ag} and include the mass of the charged lepton in the final state. The cross section $\sigma ( \pi^+ p \rightarrow \pi^+ p)$ is the production of hadrons at the center of mass energy $W$ for the reaction $\nu_\mu p \rightarrow \mu^- p \pi^+$ and is taken from~\cite{Beringer:1900zz}. 
In the present calculation the $\pi^+ p \to \pi^+p$ cross sections are evaluated at $W$ instead of $\nu$ and this modifies slightly the curves in figure 1. A second improvement is the appearance of the additional terms proportional to $\tilde{L}_{l0}$ and $\tilde{L}_{ll}$.
For each reaction we use the appropriate experimental data. For instance, for the reaction $\nu_\mu n \rightarrow \mu^- p \pi^0$ we use the reaction $\sigma ( \pi^+ n \rightarrow \pi^0 p ) = \sigma (\pi^- p \rightarrow \pi^0 n)$ from~\cite{Sadler:2004yq} as we shall discuss below. The fact that we use data means that the nonresonant background is already included.

\begin{figure}
	\centering
		\includegraphics{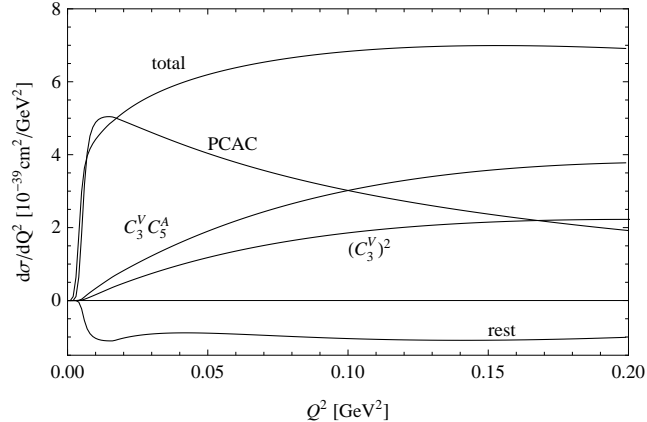}
	\caption{Anatomy of the various contributions to the cross section at $E_\nu = \unit[1]{GeV}$.}
	\label{fig:1}
\end{figure}

For the vector and interfererence terms we use the formalism of reference~\cite{Lalakulich:2005cs,Paschos:2003qr}, where the vector form factors where determined for the $I=\nicefrac{3}{2}$ amplitudes. Precise electroproducrion data established a small $I=\nicefrac{1}{2}$ nonresonant amplitude which we also include later by augmenting $C_3^V$ by 5\%[10]. 
For the vector and interference terms we use the formalism of reference~\cite{Lalakulich:2005cs} with the vector form factors from~\cite{Paschos:2011ye}. For the vector-axial interference $C_5^A C_3^V$ we use the form factor $C_5^A (Q^2)$ extracted through PCAC, $C_4^A = - \frac{1}{4} C_5^A$ and the vector form factors just described. The second column in table~\ref{tab:ISOreactions} gives the sign of the interference term which is constructive for neutrinos and destructive for antineutrinos. With these inputs we calculated the various contributions to the reaction $\nu_\mu p \rightarrow \mu^- p \pi^+$ shown in figure~\ref{fig:1}. The curve denoted as \emph{rest} indicates smaller contributions from additional form factors beyond $C_3^V$ and $C_5^A$.

\begin{table}
\begin{tabular}{|c|c|c|c|c|c|}
\hline 
reaction		& sign of $\mathcal{W}_3$	& lepton mass	& CGC				& $C_i^A\times$	& $C_i^V\times$\\
\hline
$\nu_\mu p \rightarrow \mu^- X^{++}$					& $+$	& $m_\mu$	& 1				& 1	& 1 \\
$\overline{\nu}_\mu p \rightarrow \mu^+ X^0$				& $-$	& $m_\mu$	& $\nicefrac{1}{\sqrt{3}}$	& 1	& 1 \\
$\nu_\mu n \rightarrow \mu^- X^{+}$					& $+$	& $m_\mu$	& $\nicefrac{1}{\sqrt{3}}$	& 1	& 1 \\
$\overline{\nu}_\mu n \rightarrow \mu^+ X^-$				& $-$	& $m_\mu$	& 1				& 1	& 1 \\
$\nu_\mu p \rightarrow \nu_\mu X^+$					& $+$	& $0$		& $\nicefrac{1}{\sqrt{3}}$	& y	& x \\
$\overline{\nu}_\mu p \rightarrow \overline{\nu}_\mu X^+$		& $-$	& $0$		& $\nicefrac{1}{\sqrt{3}}$	& y	& x \\
$\nu_\mu n \rightarrow \nu_\mu X^0$					& $+$	& $0$		& $\nicefrac{1}{\sqrt{3}}$	& y	& x \\
$\overline{\nu}_\mu n \rightarrow \overline{\nu}_\mu X^0$		& $-$	& $0$		& $\nicefrac{1}{\sqrt{3}}$	& y	& x \\
\hline
\end{tabular}
\caption{Input quantities and isospin factors for various reactions.}
\label{tab:ISOreactions}
\end{table}

\begin{figure}
\subfigure[$\nu_\mu p \rightarrow \mu^- p \pi^+$]{\label{fig:2a}\includegraphics[width=0.49\textwidth]{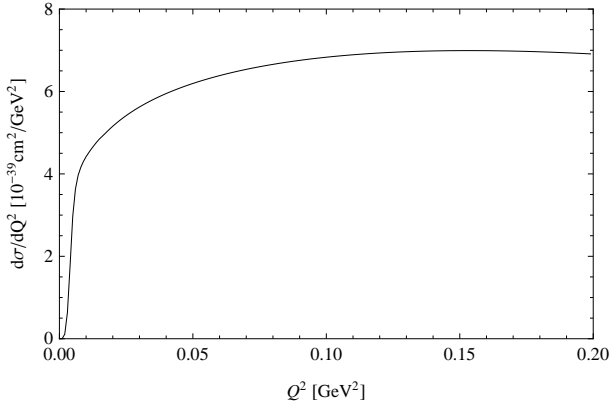}}\hfill
\subfigure[$\overline{\nu}_\mu p \rightarrow \mu^+ n \pi^0$]{\label{fig:2b}\includegraphics[width=0.49\textwidth]{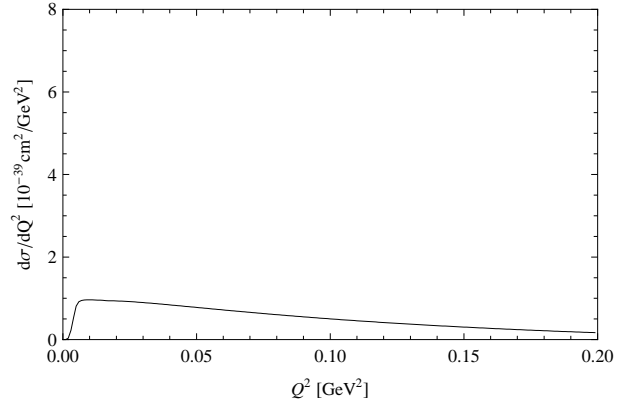}} \\
\subfigure[$\nu_\mu n \rightarrow \mu^- p \pi^0$]{\label{fig:2c}\includegraphics[width=0.49\textwidth]{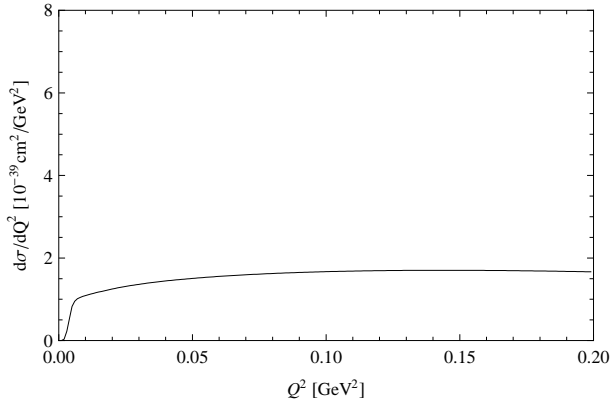}}\hfill
\subfigure[$\overline{\nu}_\mu n \rightarrow \mu^+ n \pi^-$]{\label{fig:2d}\includegraphics[width=0.49\textwidth]{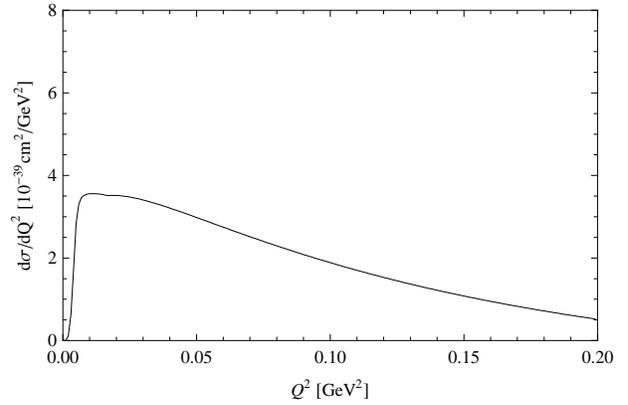}} \\
\caption{Charged current differential cross sections for $E_\nu = \unit[1]{GeV}$.}
\label{fig:2}
\end{figure}

We repeated the computation for the reaction $\nu_\mu n \rightarrow \mu^- p \pi^0$ using data from~\cite{Sadler:2004yq}. We shall use both cross sections in the comparison with the MiniBooNE data~\cite{AguilarArevalo:2010bm,AguilarArevalo:2010xt}. The results are shown in figures~\ref{fig:2a} and~\ref{fig:2c}. For antineutrino reactions the sign of the $\mathcal{W}_3 (Q^2,\nu)$ changes as indicated in table~\ref{tab:ISOreactions}. The antineutrino differential cross sections are shown in figures~\ref{fig:2b} and~\ref{fig:2d}.

\begin{figure}
\subfigure[$\nu_\ell p \rightarrow \ell^- p \pi^+$]{\label{fig:3a}\includegraphics[width=0.49\textwidth]{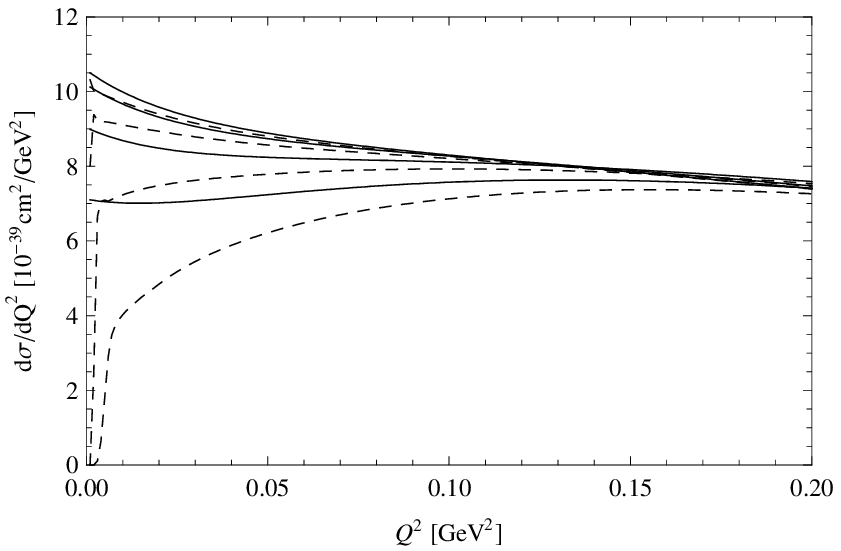}}\hfill
\subfigure[$\overline{\nu}_\ell n \rightarrow \ell^+ n \pi^-$]{\label{fig:3b}\includegraphics[width=0.49\textwidth]{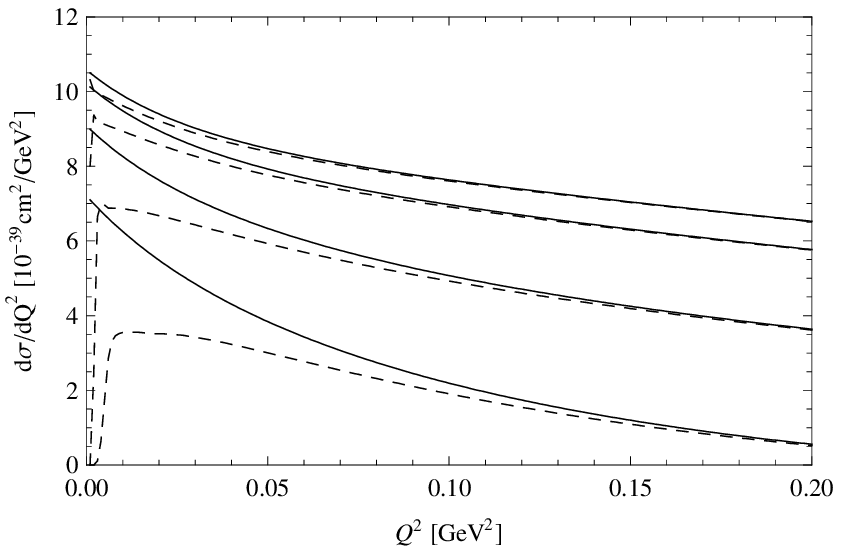}}
\caption{Comparison of muon and electron neutrino or antineutrino cross section for various energies $E_{\nu,\overline{\nu}}=\unit[1, 2, 5, 10]{GeV}$ (bottom to top). The solid line is for electron neutrinos or antineutrinos and the dashed one for muon neutrinos and antineutrinos, respectively.}
\label{fig:3}
\end{figure}

For an electron-neutrino beam we must introduce the very small mass of the electron. The result for electron- and muon-neutrino cross sections are shown in figure~\ref{fig:3} for various incident energies. For comparison we include in the same figures the cross sections for $\nu_\mu$ beams. The muon mass turns the cross sections to zero as $Q^2\rightarrow 0$. The mass effect for the $\nu_e$s is invisible.

For neutral current reactions there are more changes. The effective interaction is
\begin{align}
\mathcal{H}_{\textmd{eff}} = \frac{G_F^2}{\sqrt{2}} \overline{\nu} \gamma^\mu \left( 1 - \gamma_5 \right) \nu \left[ x \mathcal{V}_\mu^3 + y \mathcal{A}^3_\mu + \gamma \mathcal{V}_\mu^0 \right]
\end{align}
with $\mathcal{V}_\mu^3$ and $\mathcal{A}^3_\mu$ the isovector and $\mathcal{V}_\mu^0$ the isoscalar hadronic currents. The parameters in the hadronic current are given in terms of the weak angle $\theta_W$
\begin{align}
x = 1 - 2 \sin^2\theta_W, \hspace{6mm} y = -1 \hspace{6mm} \textmd{and} \hspace{6mm} \gamma = - \frac{2}{3} \sin^2\theta_W
\end{align}
with $\sin^2\theta_W \approx 0.25$. The value of $y=-1$ gives a constructive $\mathcal{W}_3$ interference term (because of the structure of the lepton current $\overline{\nu} \gamma^\mu \left( 1 - \gamma_5 \right) \nu$), making the neutrino reaction larger than the antineutrino. 
Beyond these parameters there is an overall normalization factor in the amplitudes. In the charged current interaction appears the current
\begin{align}
 \mathcal{A}_\mu^1 + i \mathcal{A}_\mu^2 = \sqrt{2} \left( \frac{\mathcal{A}_\mu^1 + i \mathcal{A}_\mu^2}{\sqrt{2}} \right) = \sqrt{2} \mathcal{A}_\mu^+
\end{align}
and for the neutral current $\mathcal{A}_\mu^3$. The CGCs are valid for the triplet $\left(\mathcal{A}_\mu^+, \mathcal{A}_\mu^3, \mathcal{A}_\mu^- \right)$. Since we have taken the amplitude for the reaction $\nu_\mu \rightarrow \mu^- \Delta^{++}$ as the standard amplitude, we divided the neutral current CGCs in the fourth column of table~\ref{tab:ISOreactions} by $\sqrt{2}$.

\begin{figure}
\subfigure[$\nu_\mu p (n) \rightarrow \nu_\mu n \pi^+ (p\pi^-)$ (solid) \newline $\nu_\mu  p (n) \rightarrow \nu_\mu  p \pi^0 ( n \pi^0 )$ (dashed)]{\label{fig:4a}\includegraphics[width=0.49\textwidth]{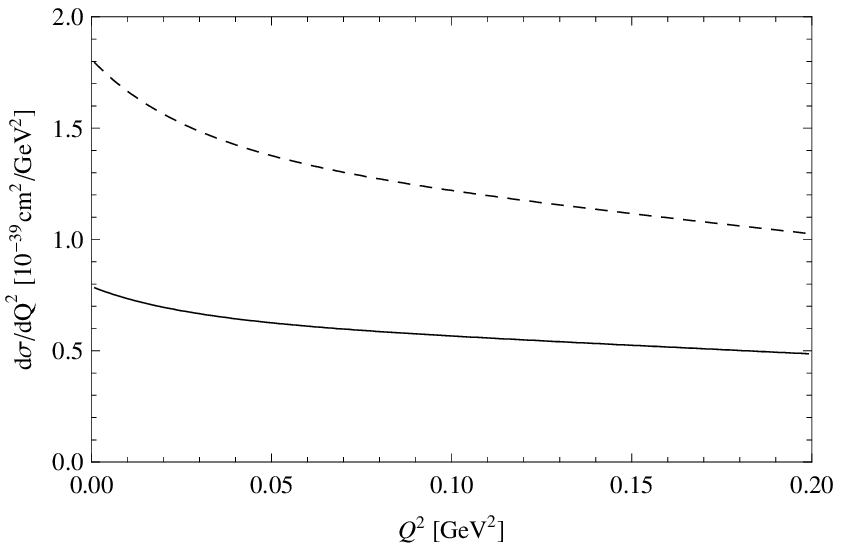}}\hfill
\subfigure[$\overline{\nu}_\mu p (n) \rightarrow \overline{\nu}_\mu n \pi^+ (p\pi^-)$ (solid) \newline $\overline{\nu}_\mu  p (n) \rightarrow \overline{\nu}_\mu  p \pi^0 ( n \pi^0 )$ (dashed)]{\label{fig:4b}\includegraphics[width=0.49\textwidth]{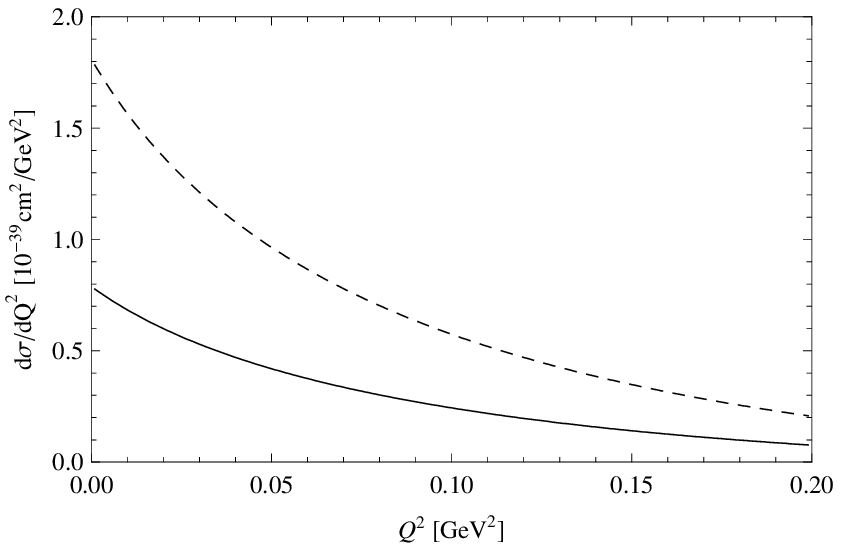}}
\caption{Neutral current reactions at $E_\nu=\unit[1]{GeV}$.}
\label{fig:4}
\end{figure}

An additional property of neutral current reactions is
\begin{align}
\sigma (\nu p \rightarrow \nu \Delta^+ ) & = \sigma (\nu n \rightarrow \nu \Delta^0 ) \label{eq:5}\\
\sigma (\overline{\nu} p \rightarrow \overline{\nu} \Delta^+ ) & = \sigma (\overline{\nu} n \rightarrow \overline{\nu} \Delta^0 ),\label{eq:6}
\end{align}
which follows from isospin symmetry which for neutral current interactions is broken by the small term $\gamma V_\mu^0$. The calculated differential cross sections for neutral currents are presented in figure~\ref{fig:4}. The equalities in equations~(\ref{eq:5}) and~(\ref{eq:6}) were also confirmed by the results of an analytical calculation in table~2 of reference~\cite{Hernandez:2007qq}. The zero mass of the neutrino produces the finite values for the cross sections at $Q^2=0$. This is also the exact point determined by PCAC where the neutrino and antineutrino cross sections are equal.

\section{Comparison with recent experiments}
\label{sec:ComparisonWithRecentExperiments}
In the resonance region pions are produced by charged and neutral currents. For our work the neutral current reactions are not relevant because the experiments do not determine the momentum transfer squared, since the momentum of the final neutrino is not observable. They measure the momentum spectrum of the produced pion, which we cannot calculate since it involves the integral over small and large values of $Q^2$. For this reason we calculate the charged current differential cross section for $\pi^+$ and $\pi^0$ production and compare them with data.

In the previous section we described the initial interactions on free protons and neutrons. In the nucleus the production and development of the resonance and of the decay products is influenced by the medium. The effects of the medium are included in a transport matrix~\cite{Adler:1974qu,Paschos:2004qh} whose nature is determined by an absorption term and the interactions of the pions. For this purpose we introduce an inverse interaction length $\kappa$ and a charge exchange matrix.

The inverse interaction length is assumed to be the same for all pions, given by 
\begin{align}
\kappa & = \rho (0) \sigma_{\textmd{tot}} (W) \\
\sigma_{\textmd{tot}} (W) & = \sigma_{\textmd{abs}} + \frac{1}{3} \sigma_{\pi^+ p} (W) \left[ h_+ (W) + h_- (W) \right]
\end{align}
with $\rho(0)$ the nuclear density at the center of the nucleus and $h_\pm (W)$ describes the forward- and backward-hemisphere projections of the Pauli blocking factor and $\sigma_{\textmd{abs}}$ is the absorption cross section given in equation~(27) of~\cite{Adler:1974qu}; it is a phenomenological function which contains several effects: the excitation of the nucleus, the propagation of the delta resonance before decay etc. Some authors compute the interaction of the delta with the medium as a self-energy correction~\cite{Oset:1987re} which shifts the mass and width of the resonance. Introducing a Breit-Wigner form for the resonance (with a corrected width $\Gamma+\delta\Gamma$) and taking its Fourier transform it becomes evident that a shift in the width corresponds to a shift in the mean free path
\begin{align}
\int_{-\infty}^\infty 
\frac{\d W \e^{iWt}}{(W-M)^2+\frac{(\Gamma+\delta\Gamma)^2}{4}} 
=\frac{2\pi}{\Gamma} \e^{iMt} \e^{-\frac{\Gamma+\delta\Gamma}{2} \frac{z}{v}}
\end{align}
with $z=vt$ the length of propagation and $v$ the velocity. This shift in the width of a resonance in momentum space corresponds to a shift in the inverse interaction length in configuration space. Thus using $\sigma_{abs}$ accounts for several effects.

During the propagation, scatterings take place as described by the pion-nucleon cross sections at the center of mass energy of the delta resonance. The final result $M(_6\textmd{C}^{12})$ has a simple form with an effective absorption factor $A(Q^2)$ and a charge exchange matrix.
For the carbon target~\cite{Paschos:2004qh}
\begin{align}
M(_6\textmd{C}^{12}) = A(Q^2) 
\begin{pmatrix} 
0.83	&	0.14	&	0.04	\\
0.14	&	0.73	&	0.14	\\
0.04	&	0.14	&	0.83
\end{pmatrix}
\label{eq:9}\end{align}
with
\begin{align}
A(Q^2 = \unit[0.05]{GeV^2}) & = 0.71 \\
A(Q^2 = \unit[0.20]{GeV^2}) & = 0.79 \\
A(Q^2 = \unit[0.40]{GeV^2}) & = 0.81 
.\end{align}
The pion multiple-scattering is solved analytically as a stochastic problem~\cite{Adler:1974qu}. The results of calculations confirm an old suggestion that charge-exchange corrections are substantial. 
They also follow a general principle: In a lepton-nucleus interaction the pions which have the same charge as the current are reduced. For pions with different charge than that of the exchanged current, the cross section is enhanced.

We apply this formalism to the MiniBooNE results. The target in the experiment is the molecule CH$_2$, which we consider as the incoherent sum of C$^{12}$ and two protons. This is justified since the two structures are relatively apart in the molecule.
The final yields of $\pi^+$ and $\pi^0$ are indicated by $\Sigma_{\pi^+}^f$ and $\Sigma_{\pi^0}^f$ and are obtained from
\begin{align}
\Sigma_{\pi^+}^f & = A(Q^2) \left\{ 0.83 \left[ \Sigma_{\pi^+}^p + \Sigma_{\pi^+}^n \right] + 0.14 \Sigma_{\pi^0}^n \right\} 6 + 2 \Sigma_{\pi^+}^p \\
\Sigma_{\pi^0}^f & = A(Q^2) \left\{ 0.73 \Sigma_{\pi^0}^n + 0.14 \left[ \Sigma_{\pi^+}^p + \Sigma_{\pi^+}^n \right] \right\} 6
.\end{align}
The cross sections within the brackets are defined as
\begin{align}
\Sigma_{\pi^+}^p & = \frac{\d\overline{\sigma}}{\d Q^2} \left( \nu_\mu p \rightarrow \mu^- p \pi^+ \right) \\
\Sigma_{\pi^+}^n & = \frac{\d\overline{\sigma}}{\d Q^2} \left( \nu_\mu n \rightarrow \mu^- n \pi^+ \right) \\
\Sigma_{\pi^0}^n & = \frac{\d\overline{\sigma}}{\d Q^2} \left( \nu_\mu n \rightarrow \mu^- p \pi^0 \right)
\end{align}
which we call the initial or primitive cross sections. The bars over the cross sections indicate averaging over the flux for $\unit[0.50]{GeV} \leq E_\nu \leq \unit[2.00]{GeV}$ and are the cross sections we computed for the spectrum of the MiniBooNE experiment.

\begin{figure}
\subfigure[$\pi^+$ yield]{\label{fig:5a}\includegraphics[width=0.49\textwidth]{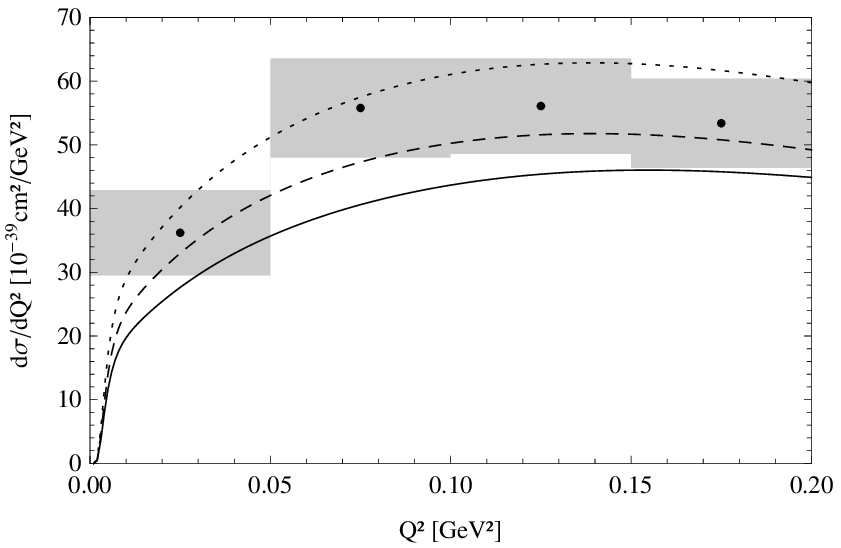}}\hfill
\subfigure[$\pi^0$ yield]{\label{fig:5c}\includegraphics[width=0.49\textwidth]{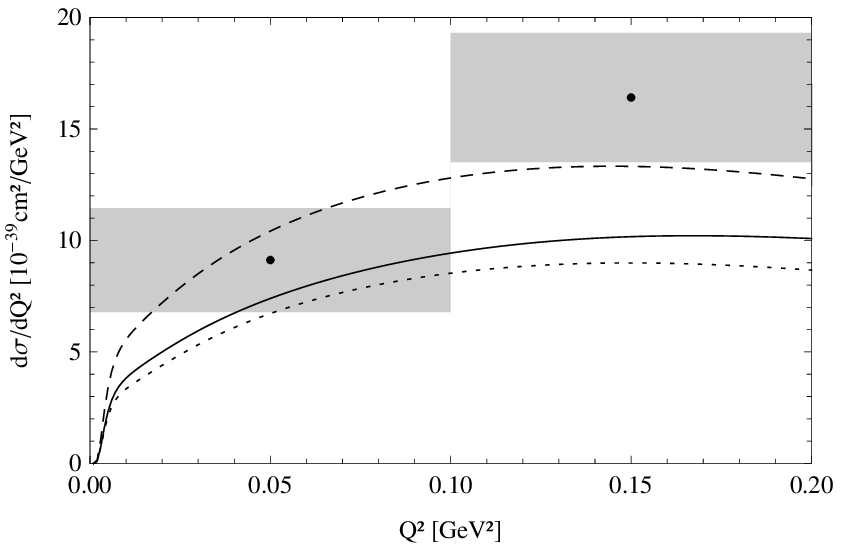}}
\caption{Charged and neutral pion production for the MiniBooNE spectrum and comparison with experimental data. The dotted line is the cross section without nuclear corrections, the dashed line with constant absorption $A=0.9$ and the solid line with $A(Q^2)$ interpolated from equations~(10) to~(12).}
\label{fig:5}
\end{figure}

The final yields for $\pi^+$ and $\pi^0$ are shown in figure~\ref{fig:5} where we plotted also the data from MiniBooNE. As the vector contribution from $C_3^V$ does not include nonresonant background~\cite{Paschos:2003qr} we augmented this form factor by 5\% as a conservative estimate in this calculation. For the effective absorption factor $A(Q^2)$ we use the values from equations~(11)--(13). The agreement of the $\pi^+$-yield with the data is very good. For the $\pi^0$-yield our curves are a little lower. However, in view of the larger experimental errors and the simplicity of the model the agreement is satisfactory.

On the experimental side, the data, especially the $\pi^0$-yield, still have large error bars and on the theoretical side it may be necessary to add a background to the $\nu_\mu n \rightarrow \mu^- \Delta^+$ production. Looking into the composition of the yields, one finds the $\pi^+$ yield comes primarily from the direct production of $\pi^+$s with a small feeding from the $\pi^0$s. The $\pi^0$ yield, on the other hand, receives a substantial feeding from the primitive $\pi^+$. The contributions from the two terms, direct and feeding, are almost equal in this channel, as the reader can easily verify using the monoenergetic initial cross sections in figure~\ref{fig:2} and the equations in this section. However, as we mentioned above, for the curves in figure~\ref{fig:5} we averaged over the neutrino spectrum. 
The contribution from charge exchange (feeding from the $\pi^+$ yield) compensates the reduction from $A(Q^2)$ and produces the solid curve in fligure~\ref{fig:5}.

\section{Conclusions}
\label{sec:Conclusions}
\begin{enumerate}
\item 
We presented a calculation for the production of pions by neutrinos in the small $Q^2$ region and for low energies ($0.50$ -- $\unit[2.00]{GeV}$), based on PCAC, CVC and experimental data. 
We selected this region because the method is based on tested physical principles.
The results agree with the Argonne and Brookhaven data for the $\pi^+$ channel, as has been shown earlier~\cite{Paschos:2011ye}. The results are consistent with a value for $C_5^A (Q^2)$ close to $1.1$ and a small background. A small background is also observed in electro-production~\cite{Galster:1972rh} by comparing the $\pi^0$ and $\pi^+$ channels. A small background was also obtained in calculations at the static limit~\cite{Adler:1968tw,Campbell:1973wg}. The same Argonne experiment requires a large $I=\nicefrac{1}{2}$ component, however one must keep in mind, that the statistics for the reaction $\nu_\mu n \rightarrow \mu^- \Delta^+$ are limited.

\item
We applied our method to the MiniBooNE experiment, which observed two different final states ($1\pi^+$ and $1\pi^0$), and found good agreement, which further supports our estimates of the initial interactions and in addition for the transport model for the carbon nucleus~\cite{Paschos:2004qh}. For these results we used flux averaged cross sections. The $\pi^+$ cross section is affected by absorption and Fermi blocking with a very small correction from charge exchange. In contrast the $\pi^0$ channel receives a 45\% contribution from charge exchange as the large $\pi^+ p$ channel feeds into it. This is very different from the results of GiBUU~\cite{Mosel:2012kt} where they reported (see end of section~III.~B in~\cite{Lalakulich:2012cj}) that the ``FSI have a relatively small effect''  on the $\pi^0$ yield by reducing it, instead of increasing it.

\item
One of the uncertainties in neutrino reactions is the presence of many form factors, whose $Q^2$ dependence is, in principle, unknown and determined from data. One can use for the leading form factor, especially for $C_5^A(Q^2)$, the functional form we found and match it to the extrapolation to larger vales of $Q^2$ (see figure~3 in~\cite{Paschos:2011ye}). Thus we consider the results at $Q^2\leq \unit[0.20]{GeV^2}$ as a starting point for the parameterizations to larger values.

\item
The charge exchange matrix in equation~(\ref{eq:9}) and the primary interactions obey isospin symmetry. Thus for isoscalar targets, like carbon, there are isospin relations between various reactions which can be tested in the new experiments (Minerva). In an internal report~\cite{Paschos:2012va} we elaborated on some of them. They should be tested in the experiments. Finally, the input parameters for the initial reactions~\cite{Paschos:2011ye} and the nuclear corrections~\cite{Paschos:2004qh} were published earlier and we did not need to modify them.
\end{enumerate}


\end{document}